\magnification 1100
\hsize=6.5truein
\vsize=8.5truein
\baselineskip=16pt

\input epsf

\def \vec#1{ {\bf #1} }
\def \state#1{ {\cal #1} }
\def \gcc{\ifmmode {\rm g~cm^{-3}} \else g~cm$^{-3}$ \fi}  
\def \lmin{ l_{\rm min} }
\def \lmax{ l_{\rm max} }

\centerline{\bf FULLY THREADED TREE}
\centerline{\bf FOR ADAPTIVE REFINEMENT FLUID DYNAMICS SIMULATIONS}

\vskip 2 cm

\centerline{A.M. Khokhlov}
\bigskip\centerline{Laboratory for Computational Physics and Fluid Dynamics,}
        \centerline{Code 6404, Naval Research Laboratory,}
        \centerline{Washington, DC 220375}

\vfill\eject  

\centerline{\bf Abstract}

\bigskip
A fully threaded tree (FTT) for adaptive refinement of regular meshes
is described. 
By using a tree threaded at all levels, tree traversals for
finding nearest neighbors are avoided.
All operations on a tree including tree modifications are ${\cal O}(N)$, 
where $N$ is a number of cells, and are performed in parallel.
An efficient implementation of the tree is described that requires 
$2N$ words of memory.
A filtering algorithm for removing high-frequency noise during
mesh refinement is described.

A FTT can be used in various numerical applications.
In this paper, it is applied to the integration of the Euler equations of 
fluid dynamics.
An adaptive-mesh time stepping algorithm is described in which
different time steps are used at different levels of the tree.
 Time stepping and mesh refinement are interleaved
to avoid extensive buffer layers of fine mesh which were otherwise required
 ahead of moving shocks.
Test examples are presented, and the FTT performance is evaluated.
The three-dimensional simulation of the interaction of a shock wave and a
spherical bubble  is carried out that shows the development of azimuthal 
perturbations on the  bubble surface.
\vfill\eject

\centerline{\bf 1. Introduction}

\bigskip
Adaptive mesh refinement (AMR) is used to increase spatial and temporal
resolution of numerical simulations beyond the limits imposed by the 
available hardware. In fluid dynamics, AMR is used in
simulations of both steady and unsteady flows on structured and 
unstructured meshes [1-23]. A mesh is {\sl structured}
if it is composed of cells of rectangular shape, otherwise it is
called unstructured. Integration of the equations of fluid dynamics 
is easier and the overhead in computer time and memory per computational cell
are minimal for structured meshes. The corresponding overhead is
usually high for  unstructured meshes.  However, an unstructured mesh has
the advantage that it can  conform well
to a complex boundary. Whether to use a 
structured or unstructured approach depends on the specific problem,
 code availability, and general preferences.
This paper deals with structured meshes, and specifically with 
a structured-mesh AMR.

There are two different approaches to structured--mesh AMR.
In one standard approach, cells are organized as two-dimensional or
three-dimensional arrays (grids). A coarse base grid represents
 the entire computational domain. Finer, nested grids are layed over
coarser grids where more resolution is required. A grid hierarchy may be
organized as a tree  [1--12]. 
One advantage of this approach is that any single-grid fluid flow solver can 
be used without modifications for AMR. However,
grids themselves are inflexible structures. It 
is difficult to cover complex features of a flow with  a few 
grids. Many grids are typically required, and a substantial
number of computational cells can be wasted on a smooth flow.
Several grids of the same level of refinment may overlap
which leads to duplication of cells.
Periodic rebuilding of the entire grid hierarchy is required when the 
flow evolves with time. 

In an alternative approach, individual 
 computational cells are organized directly in a tree [13].
Each cell can be refined or unrefined separately from the others, as needed.
At every level of the tree, the mesh may have an arbitrary shape, 
as opposed to grids. Duplication of cells is avoided.
To date, this approach has been used mainly in steady-state fluid flow 
simulations [13--20]. 
It has been applied to unsteady flows in two dimensions [21--23].
The main advantages of the tree approach are the flexibility in refinment and 
unrefinement, and the efficiency in using cells. 
However, there are certain problems with the tree approach:
(1) Access to neighboring cells is more difficult in a tree
than in an array. Tree traversals to access nearest
neighbors are difficult to vectorize and parallelize. 
(2) The memory overhead to maintain a tree, though less than for 
unstructured meshes, is still substantial [21-23]. Thus, 
an efficient parallel
access to information on a tree, and reduction of the memory overhead
are the two problems that need to be solved.

Another problem, (3),  arises in unsteady
flow simulations with many levels of tree refinement. 
In unsteady simulations, buffer layers of finely 
refined cells must be created in advance and ahead of shocks to prevent them
from running out of the fine mesh.
Let the maximum and minimum levels of mesh refinement be
$\lmax$ and $\lmin$, respectively, and the cell refinement ratio be a factor
of two. If refinement is done between time steps [21-23], the
 thickness of the buffer layers should be at least
 $2^{\lmax-\lmin}$ finest cells. The number of required cells
then increases exponentially when 
$\lmin-\lmax$ increases, and becomes
prohibitively large, especially in three dimensions,
  if $\lmax-\lmin$ is large.

This paper addresses the problems outlined above.
To alleviate  problems (1)
and (2), a new structure, a {\sl fully threaded tree} (FTT),
is designed. In an ordinary tree, each cell has pointers 
to its children, if any. The memory for keeping these pointers is not
used in leaves (cells which do not have children).
In a threaded tree, this free memory is used to organize ``threads'' 
along which the tree can be quickly traversed during various
search operations. A discussion of a binary threaded tree
in which threads are used for left-to-right and right-to-left search 
can be found in [24].
A binary-, quad-, or an oct-tree is required to organize a one-, two-,
or three-dimensional regular mesh, respectively. 
An implementation of an oct-tree would require at least 8 words of memory
per cell. Some of this memory can then be used to make threads through
leaves in order to facilitate finding the nearest neighbors.
However, this scheme cannot find neighbors of split cells, and 
cannot be parallelized (Section 4).
In the FTT described in this paper, every cell,
 whether a leaf or not, has an easy access to its 
children, neighbors and parents. One may say that an FTT is a tree
 threaded in all possible directions.  At the same time, 
the memory overhead  for supporting an FTT is significantly reduced. 
The maintenance of an octal FTT requires two words of memory per 
computational cell only.
Another important property of a FTT is that 
all operations on it, including modifications of the tree structure itself,
can be performed in parallel. This then allows
vectorization and parallelization of tree-based AMR algorithms.

In this paper, the FTT is used for the integration of the Euler equations of 
fluid dynamics. To alleviate the third problem outlined above,
the integration in time and tree refinement
have been  coupled together,
and  time stepping at different levels of the tree and 
tree refinements of these levels have been interleaved. As a result, 
excessive buffer layers of refinement
ahead of moving shocks are no longer needed.

The following sections describe the equations solved (Section 2),
tree structure and discretization of the equations on the tree (Section 3),
tree implementation (Section 4),
integration of the Euler equations (Section 5),
refinement and unrefinement (Section 6), numerical tests (Section 7), and
FTT performance (Section 8).
\vfill\eject
\bigskip\centerline{\bf 2. Equations}

\bigskip
In this paper, the FTT is used to solve the Euler equations for an inviscid
flow
$$
\eqalign{
   &  {\partial\rho\over\partial t}~+~\nabla\cdot(\rho\vec{U})~=~0~, \cr
   &  {\partial \rho \vec{U} \over \partial t}
      ~+~\nabla\cdot(\rho\vec{U}\vec{U})+\nabla P~=\rho\vec{g}~, \cr
   &  {\partial E \over \partial t}~+~\nabla\cdot((E+P)\vec{U})~=~
             \rho\vec{U}\cdot\vec{g}~,
                                                       \cr} \eqno(1)
$$
where $\rho$ is the mass density, $\vec{U}$ is the fluid velocity, 
$E$ is the total energy density, $P=P(E_i,\rho)$  is the pressure, 
$E_i = E - \rho U^2/2$ is the internal energy density,
and $ \vec{g}$  is an external acceleration.

Numerical integration of (1) requires  evaluating of
numerical fluxes of  mass, momenta, and energy at cell interfaces.
We use a second order accurate Godunov-type method 
based on the solution of a Riemann problem to evaluate the fluxes
[26-28]. This is one of the monotone methods known to
provide good results for both flows with strong
shocks, and for moderately subsonic and turbulent flows [32-34].
 However, the
algorithms described in this paper are general, and can be
used in cojunction with other high-order monotone methods such as
the Flux-Corrected-Transport [35], for example.

\bigskip\centerline{\bf 3. FTT structure and discretization}

\bigskip
A computational domain is a cube of size $L$. It is subdivided to
a number of cubic cells of various sizes $1/2,~1/4,~1/8,...$ of $L$.
Cells are organized in an oct-tree with the entire computational domain 
being the root. For the integration and refinement
algorithms to be effective, the following information must be 
easily accessible for every cell $i$:
\smallskip
\settabs\+\noindent&*****&***********&******&\cr
\+&& $iLv(i)$            &-& level of the cell in the tree \cr
\+&& $iKy(i)$            &-& TRUE/FALSE if cell is split/unsplit \cr
\+&& $iPr(i)$            &-& pointer to a parent cell \cr
\+&& $iCh(i,j)$          &-& pointers to children, $ j = 1 \div 8$ \cr
\+&& $iNb(i,j)$          &-& pointers to neighbors, $j =1\div 6$. \cr
\smallskip\noindent
The cell size is related  to $iLv(i)$ by $\Delta_i = 2^{-iLv(i)} \, L$.

Cells in the tree are either split (have children) or leaves
(do not have children). 
Relations  between cells in the tree, and  the
directions of various pointers are illustrated in Figure 1 for a 
one-dimensional binary tree. Relations for a quad- or an oct-tree
are similar. In Figure 1, the root (cell 1) 
represents the entire computational
domain. It has two children (cells 2 and 3), each representing half of the 
domain. There will be four or eight children in 2-D or 3-D,
respectively.
Cell 2 is further subdivided on two cells (cells 4 and 5).
Neighboring leaves are not allowed to differ in size by more than a factor 
of two. The neighbor--neighbor relation is not reciprocal for
leaves of different size that face each other. In Figure 1,
cell 5 has cell 3 as its neighbor, but cell 3 has cell 2 as its neighbor,
and not cell 5. A $j$-th neighbor of a cell $i$
 either has the same size as the cell 
itself,  $\Delta_{iNb(i,j)} = \Delta_i$, or it is two times larger,
$\Delta_{iNb(i,j)} = 2 \Delta_i$. In the former case, the neighbor may be
a leaf or a split cell. In the latter case, it can only be a leaf.

Physical state information $\state{U}_i = \{\rho_i, E_i,(\rho \vec{U})_i\}$
 is kept in both split and unsplit cells. 
Information kept in split cells is needed 
 to evaluate mass, momenta
and energy fluxes across interfaces between leaves of different size 
(Section 5), and to make
decisions about refinement and unrefinement (Section 6).
The physical state vectors for split cells are updated by averaging over
children, when required. The memory overhead for 
keeping state vectors of split cells is 
$$
{\rm Overhead}~=~{{\rm Number~of~splits}\over{\rm Number~of~leaves}}~=~
{1\over 2^{\rm dim}} + {1\over 2^{2\rm dim}} + {1\over 2^{3\rm dim}}
 +... \simeq {1\over 2^{\rm dim} - 1}~,  
$$
where ${\rm dim}=1\div 3$ is a number of spatial dimensions. The overhead is
$\simeq 100$\% in 1-D, $\simeq  33$\% in 2-D, and $\simeq  14$\% in
3-D.

Coordinates 
$\vec{r}_i = \{x_i, y_i,z_i\}$ of cell centers are
associated with every cell.
It is assumed that neighbors 1 to 6 of a cell are left X, right X,
left Y, right Y, left Z, and right Z neighbors, respectively, and that
coordinates increase from left to right: $x_{iNb(i,1)} < x_i < x_{iNb(i,2)}$,
$y_{iNb(i,3)} < y_i < y_{iNb(i,4)}$, $z_{iNb(i,5)} < z_i < z_{iNb(i,6)}$.
Keeping coordinates is not necessary for finite-difference operations.
They are useful, however, for evaluating long-range forces (e.g., gravity) by
direct summation or multipole expansion algorithms.


\bigskip\centerline{\bf 4. Implementation of FTT}

\bigskip
The FTT, as  described above,  can be implemented 
by storing all pointers, coordinates and flags in computer memory,
and then modifying them when the tree is refined or unrefined.
However, such a direct implementation requires $20N$ words of memory,
 where $N$ is the number of
cells in the tree. This memory overhead is too large.
In addition, such a tree cannot be modified in parallel. 
Because neighbors of a cell keep pointers to that cell, cell removal
requires retargeting these pointers. Removing one cell thus affects others. 
A conflict appears if a cell and its neighbors
are to be removed simultaneously. Similar problems arise if
neighboring cells are created simultaneously. 

An improved parallel version of the tree is based on the following three
observations: ({\it i}) When a cell is split, its eight children are 
created simultaneously. Thus, they may be stored in memory contiguously [13],
 so that only one pointer is needed to find all eight
children. ({\it ii}) Since all eight children (siblings) are kept in memory 
together, neighbor--neighbor relations between them are known automatically.
There is no need for pointers between the siblings. ({\it iii}) 
Neighbors of any cell are either its siblings or they are 
children of a neighboring
parent. There is a small fixed number of neighbor-neighbor relations
between siblings of neighboring parents.
 Thus, a child's neighbor that belongs to a different parent 
can be determined without search if parents
keep pointers to their own neighbors.

The FTT structure implemented is illustrated in Figure 2.
All cells are organized in groups called octs. Each oct contains eight cells.
Each cell has a physical state vector  $\state{U}$ associated with it,
and a pointer to an oct which contains its children,
if any, or a {\sl nil} pointer.  Each oct 
knows its level, $OctLv$, which is equal to the level of the oct's cells. 
Each oct has a pointer $OctPr$ to a parent cell.
Each oct has six pointers $OctNb(6)$ to {\sl parent cells of neighboring
octs}. 
This information is enough to find neighbors,
children and parents of every cell without  searching.
The memory requirement is 16 words per oct or 2 words per cell.
The memory overhead is thus significantly reduced. The price for this is
the computer-time overhead associated with computing cell pointers from oct
pointers and oct pointers from cell pointers,
 but this overhead is relatively small (Section 8).

If needed, octs may also contain coordinates $OctPos$ of their centers.
 These coordinates
are equal to coordinates of the corresponding parent cells.
Coordinates of cells that belong to an oct can be found by adding or 
subtracting $\Delta_i/2$ from
the corresponding oct's coordinates. Memory requirement for
coordinates is thus 3 words per oct or $3\over 8$ words per cell,
which increases the total memory requirement to  $2{3\over8}$ words per cell.

Figure 2 shows that FTT allows parallel
modifications at any given tree level. To unrefine a cell,
the associated oct pointed to by $OctCh$ must be destroyed, and 
$OctCh$ must be set to {\sl nil}. No other changes are required in 
neighboring octs or cells located at the same tree level.
This is because neighboring octs are accessed by parent cells located
at a different level of the tree. 
To refine a cell, a new oct must be created with all necessary pointers,
 and the corresponding
$OctCh$ pointer of the refined cell must be set to a new oct. 
Again, no other changes in 
any other octs or cells located at the same tree level are required.

All operations on the tree
can thus be expressed as a sequence of parallel operations on large groups 
of cells that belong to the same tree level.
Therefore, it is possible to introduce a
general iterator for performing operations on these cells,
 and to hide the details of parallelelization inside it.
 This greatly reduces the amount of code, and simplifies coding.
To perform a  parallel operation, each processor first gathers
 all required information in work  arrays using indirect addressing. 
Then, the CPU-intensive work is done by looping 
 over contigous blocks of memory. After that, the result is scattered back.
In this way, the code performance is virtually unaffected by the 
distribution of cells in a physical space.

\bigskip\centerline{\bf 5. Integration procedure}

\bigskip
Equations (1) are integrated in time with different
time steps $\Delta t(l)$ at different levels of the tree $l$;
$l_{\rm min} \le l \le l_{\rm max}$, where 
$l_{\rm min}$ and $l_{\rm max}$ is the minimum and maximum levels of
leaves.
A global time step $\Delta t \equiv \Delta t(l_{\rm min})$
 is determined from the CFL condition,
$$
\Delta t = cfl \,{ 2^{-l_{\rm min}}\, L \over \max_i
    \left(\max_j\left( a_i+|U_{i,j}|\right)\right)}~,  \eqno(2)
$$
where $a$ is the sound speed, $cfl < 1$ is a constant, and maximum in 
 (2)
is taken over all three directions $j=x,y,z$  and over all leaves $i$.
Time steps at various levels are
$$
\Delta t(l) =  2^{l_{\rm min}-l} \, \Delta t ~.   \eqno(3)
$$
Integration 
at different levels of the tree is interleaved with tree refinement.
Let us designate the procedure of advancing level $l$ one step $\Delta t(l)$
in time as ${\cal A}(l)$, and the procedure of tree refinement at level $l$ as
${\cal R}(l)$. 
The ${\cal A}$ procedure is described later in this section. 
The ${\cal R}$ procedure consists of refining leaves of level $l$
and unrefining split cells of level $l$ according to certain refinement
criteria. This procedure is described in Section 6. 

A global time step of integration  can be expressed as a recursion,
$$
{\sl Global~Time~Step} = {\cal S}(\lmin)~,         \eqno(4)
$$
where the procedure ${\cal S}(l)$ is a combination of advancing and refinement
procedures 
$$
{\cal S}(l) = {\cal R}(l)\cases{{\cal S}(l+1); {\cal A}(l); 
                                {\cal S}(l+1); {\cal A}^\dagger(l);
                                                    &if $l<\lmax$;\cr
                                {\cal A}(l);  {\cal A}^\dagger(l);
                                                    &if $l=\lmax$.\cr}
                                                   \eqno(5)  
$$
All procedures in  (5) are performed from left to right.
The procedures  ${\cal S}(l+1)$ and ${\cal A}(l)$ are repeated twice
when directional time-step splitting is used in ${\cal A}$ (see below).
After every ${\cal A}$ procedure, the sequence of XYZ one-dimensional
sweeps is recursively reversed. This is indicated 
 by the $\dagger$ superscript. 
Note that $\cal R$ can change both $\lmax$ and $\lmin$, recursively.
Therefore, the sequence of advances and refinements generated
changes dynamically, and generally cannot be predicted from starting values
of $\lmin$ and $\lmax$. The sequence can be determined in advance only if
$\lmin$ and $\lmax$ are fixed during a global time step.
For example, for $\lmin=6$ and $\lmax=8$ fixed,
the sequence generated  would be
                 [{$\cal R$}(6) [{$\cal R$}(7) [{$\cal R$}(8) 
                                            {$\cal A$}(8) 
                                            {$\cal A$}$^\dagger$(8)]
                              {$\cal A$}(7) 
                                         [{$\cal R$}(8) 
                                            {$\cal A$}(8) 
                                            {$\cal A$}$^\dagger$(8)]
                              {$\cal A$}$^\dagger$(7)]
               {$\cal A$}(6) 
                          [{$\cal R$}(7) 
                                         [{$\cal R$}(8) 
                                            {$\cal A$}(8) 
                                            {$\cal A$}$^\dagger$(8)]
                              {$\cal A$}(7) 
                                         [{$\cal R$}(8) 
                                            {$\cal A$}(8) 
                                            {$\cal A$}$^\dagger$(8)]
                              {$\cal A$}$^\dagger$(7)]
               {$\cal A$}$^\dagger$(6)]. 
Square brackets separate different levels of recursion.

Now we describe the advance procedure, ${\cal A}(l)$.
The solution at every level $l$ is advanced in time using
direction splitting. 
Equations for the X-sweep can be written as
$$
{\partial \state{U}\over\partial t}~=~
{\partial \state{F}(\state{U})\over\partial x}~+~\state{S}~,    \eqno(6)
$$
where the flux vector is
$\state{F} ~=~ \left(\, \rho U_x,\, (E+P)U_x,\, \rho U_x^2, \,\rho U_xU_y, \,
                       \rho U_xU_z \, \right)$,
and the source term is
$ \state{S} = \left(\, 0,\, \rho U_x g_x,\,\rho g_x,\,0,\,0\,\right)$.
The equations for the Y- and Z-sweeps are similar.

Let $\state{U}^o$ and $\state{U}^n$ be a state vector at the beginning 
and at the end of a global time step, respectively.
Let us designate right and left neighbors of cell
 $i$ in the direction of a sweep
as  $i+$ and $i-$, respectively, and let us introduce the following quantities
$$
\alpha = 2^{\lmin}\,{ \Delta t\over L}~,~~\beta = 2^{-\rm dim} \alpha~,~~{\rm and}~
\gamma(i,l) = \cases{\alpha, &if~$iLv(i+)=l$; \cr
                   \beta, &if~$iLv(i+) = l-1$~, \cr }   
$$
where ${\rm  dim}=1\div 3$ is the number of spatial dimensions.
The ${\cal A}(l)$ procedure can then  be described in a form of the
following pseudocode:
$$\eqalign{
&{\sl for}~(~{\sl leaves}~i~{\sl of~level}~l-1~)~~\{~
{\sl if}~(~i~{\sl has~split~neighbors}~)~\state{U}^n_i := \state{U}^o_i;~\} \cr
&{\sl for}~(~{\sl X,~Y,~Z~directions}~)~~\{ \cr
&~~~~{\sl for}~(~{\sl leaves}~i~{\sl of~level}~l~)~~\{ \cr
&~~~~~~~~{\sl if}~(~i+~{\sl is~a~leaf~or~boundary}~)~~\{ \cr
&~~~~~~~~~~~~{\sl Compute~fluxes}~\state{F}_{i,i+}~{\sl at~the}~(i,i+)
          ~{\sl interface}~; \cr
&~~~~~~~~~~~~\state{U}_i^n:=\state{U}^n_i  
                 -\alpha\,\state{F}_{i,i+}+\Delta t(l) \, \state{S}_i~;     \cr
&~~~~~~~~~~~~\state{U}_{i+}^n:= \state{U}_{i+}^n 
                            + \gamma\,\state{F}_{i,i+}; \cr
&~~~~~~~~\} \cr
&~~~~~~~~{\sl if}~(~i-~{\sl is~a~leaf~of~level}~l-1~{\sl or~boundary}~)~~\{ \cr
&~~~~~~~~~~~~{\sl Compute~fluxes}~\state{F}_{i-,i}~{\sl at~the}~(i-,i)
        ~{\sl interface}~;          \cr
&~~~~~~~~~~~~\state{U}_i^n:=\state{U}^n_i +\alpha\,\state{F}_{i-,i}~;\cr
&~~~~~~~~~~~~\state{U}_{i-}^n:=\state{U}_{i-}^n
                     -\beta\,\state{F}_{i-,i}; \cr
&~~~~~~~~\} \cr
&~~~~\}    \cr
&~~~~{\sl for}~(~{\sl leaves}~i~{\sl of~level}~l~)~~\{~
    {\sl if}~(~i~{\sl has~no~neighbors~at~level}~l-1~)~
                          \state{U}_{i}^o:= \state{U}_{i}^n;~~\} \cr
&\} \cr
&{\sl for}~(~{\sl split~cells}~i~{\sl of~level}~l~)~~\{~
   \state{U}_i^o := 2^{-\rm dim}\, \sum_{j=1}^{2^{\rm dim}}
                \state{U}^o_{iCh(i,j)};~\} \cr
&{\sl for}~(~{\sl leaves}~i~{\sl of~level}~l~)~~\{~
   {\sl if}~(~i~{\sl has~split~neighbors}~)~
                               \state{U}^o_i := \state{U}^n_i;~~\}~. \cr
}       \eqno(7)
$$
The factor $2^{-\rm dim}$ in $\beta$, $\gamma$, and in  (7)
 takes into account the difference in
volumes of neighboring cells of different sizes.

The algorithm (4)--(7) 
is schematically illustrated in Figure 3 for the two-dimensional case.
In this particular example, the tree has only two levels occupied by leaves,
$\lmax = \lmin+1$.
 Cells 1, 2, 3 and 4 belong to level $\lmax$,
and cells 5 and 6  belong to level $\lmin$. 
During a time step at level $l=\lmax$, fluxes accross interfaces
(1,3), (2,4), (3,5), and (4,5) are evaluated, new state vectors $\state{U}^n$
for cells 1, 2, 3 and 4 are computed, and old state vectors $\state{U}^o$
in these cells are updated. For cell 5, $\state{U}^n_5$ is changed
due to fluxes $\state{F}_{3,5}$ and $\state{F}_{4,5}$.
However,  state vector $\state{U}^o_5$ is not updated yet.
During the second time step $\Delta t(l)$, this procedure is repeated:
States $\state{U}^0_1$, $\state{U}^0_2$, $\state{U}^0_3$, $\state{U}^0_4$ 
are updated,
$\state{U}^n_5$ is changed again due to 
fluxes $\state{F}_{3,5}$ and $\state{F}_{4,5}$, but 
 state vector $\state{U}^o_5$ is still not updated.
After two time steps at level $l$, one
time step at level $l-1$ is performed.
Fluxes at  interface (5,6) are computed, 
 $\state{U}^n_5$ is changed due to these fluxes, and 
the state vector $\state{U}^o_5$ is finally updated.

For leaves that have neighbors at the same tree level,
 the procedure is that of
usual conservative finite differencing 
$$
\state{U}^n_i ~=~ \state{U}^o_i+ {\Delta t(l)\over \Delta x_i} 
\left( \state{F}_{i,i-} - \state{F}_{i,i+}\right)    \eqno(8)
$$
with left and right fluxes evaluated at the same time $t=t^o + \Delta t(l)/2$.
For leaves that have split neighbors, flux contributions from more than
 two neighbors at two different moments of time, $t=t^o + \Delta t(l)/2$ and
$t=t^o + 3\Delta t(l)/2$,  are taken into account,
again in a conservative fashion.

Fluxes are obtained by solving a Riemann problem at every interface according
 to [26]. Left and right states for the Riemann problems are obtained by a 
piecewise linear reconstruction [27] from $\state{U}^o$ state vectors.
A small amount of dissipation is introduced into the code as an
artificial diffusion  added to the numerical
fluxes [28]. 
For cells that have neighbors at the same tree level, the entire
integration procedure is formally second order accurate both in space 
and in time. For cells that have neighboring leaves at different levels, the
 accuracy  reduces to first order.

If necessary, accelerations $\vec{g}$ are computed in the 
beginning of every advance procedure.
 To maintain the second order accuracy in time for the
source term $\state{S}$, the state $\state{U}^o$ is corrected after every
advance  procedure as 
$$
 \tilde{\vec{U}} :=\vec{U} ~+~{\vec{g}^{t+\Delta t} - \vec{g}^t
\over 2} \,\Delta t~;
~~~E := E + \rho\,{\tilde{U}^2 - U^2 \over 4}~;~~~\vec{U} :=
\tilde\vec{U}~. \eqno(9)
$$

\bigskip\centerline{\bf 6. Mesh refinement}

\bigskip
The most difficult part of an AMR simulation is to
 decide where and when to refine or unrefine a mesh. 
It is also the most problem-dependent part. Different problems may require
different criteria for refinement and unrefinement. 
The refinement procedure currently implemented 
 consists of four steps:

\item{1.} For every cell, a refinement indicator,
$0 \leq \xi \leq 1$ is computed. Large $\xi > \xi_{\rm split}$ indicates that
 a leaf must be refined, and small $\xi < \xi_{\rm join}$ indicates that 
a split cell can be unrefined; $\xi_{\rm split}$ and $\xi_{\rm join}$ are
some predefined constant values.

\item{2.} $\xi$ is smoothed in order to prevent cells from being
falsely refined (mesh trashing)
in places where $\xi$ fluctuates around critical values $\xi_{\rm split}$ 
and $\xi_{\rm join}$.

\item{3.} Leaves are refined if $\xi > \xi_{\rm split}$.

\item{4.} Split cells are unrefined if $\xi < \xi_{\rm join}$
 and if they are not just split.

\smallskip
There are two  approaches to computing $\xi$. First, it may
be computed to measure the convergence  of a solution
[2]. This allows to control the accuracy
 of the solution ``on flight,'' but requires that the solutions
on the fine and coarse meshes be generated simultaneously. Another
approach is to use an indicator proportional to gradients in the solution.
Such an indicator shows where to expect a large error in the
solution [16,19,20].
The convergence of the solution must be checked afterwards with a different
resolution. We adopt the second approach, but note that the tree
structure allows both approaches to be implemented.

The indicator $\xi$ is constructed as a maximum of several
indicators, 
$$
\xi = {\rm max} ( \xi^1, \xi^2, ...)~,                    \eqno(10)
$$
each of which is either a shock indicator, contact discontinuity indicator,
 or a gradient 
indicator, all normalized to unity, $0 \leq \xi^k \leq 1$.

As a shock indicator, the following quantity is used [28]
$$\eqalign{
& \xi^s_i = \max_{j=1\div 6} \xi^s_{i,j}~,          \cr
& \xi^s_{i,j} = \cases{1, &
  if ${|P_{iNb(i,j)}-P_i|\over{\rm min}(P_{iNb(i,j)},P_i)} >\epsilon_s~
    {\rm and} ~\delta\cdot (U_{iNb(i,j),k}-U_{i,k}) > 0$; \cr
     & \cr
     0, & otherwise; \cr}                    \cr}  \eqno(11)
$$
where $\delta = 1 - 2\cdot {\rm mod}(j,2)$, $k = {\rm int}(j/2)$,
and $\epsilon_s = 0.2$ determines the minimum shock strength to be
detected. 

A discontinuity indicator is defined as follows:
$$\eqalign{
& \xi^c_i = \max_{j=1\div 6} \xi^c_{i,j}~,          \cr
& \xi^c_{i,j} = \cases{1, &
  if ${|P_{iNb(i,j)}-P_i|\over{\rm min}(P_{iNb(i,j)},P_i)} <\epsilon_s~
    {\rm and}~
{|\rho_{iNb(i,j)}-\rho_i|\over{\rm min}(\rho_{iNb(i,j)},\rho_i)} 
              >\epsilon_c~ $; \cr
     & \cr
     0, & otherwise; \cr}                    \cr}  \eqno(12)
$$
where $\epsilon_c = 0.2$.

A gradient indicator for a variable $a$ 
is constructed as
$$
\xi^a_i = \max_{j=1\div6}
        \left({||a_{iNb(i,j)}| - |a_i||\over \max(|a_{iNb(i,j)}|,|a_i|)}
\right)~,   \eqno(13)
$$
where $a$  may be a mass density, energy density, pressure, velocity, 
vorticity,  etc.

As explained in the beginning of this section, the indicator $\xi$ computed
according to  (10) -- (13) must be smoothed before it is used for
 refinement.
Smoothing
 is based on the analogy with the propagation of a reaction-diffusion 
front.
Let us consider  $\xi$ as a concentration of  a reactant obeying
a reaction--diffusion equation 
$$
{\partial \xi \over \partial \tilde{t}} = K \nabla^2 \xi + Q,   \eqno(14)
$$
where $\tilde t$ is a fiducial time, $K =2^{-2l}L^2$ 
is a constant diffusion coefficient, and
$$
Q = \cases{ 1, &if $1 > \xi > \xi_{\rm split}$; \cr
               &                        \cr
            0, & otherwise;             \cr} \eqno(15)
$$
is a reaction rate. A steady-state form of (14),
$$
 S_\xi {\partial \xi \over \partial x} = {\partial^2 \xi\over  \partial x^2} + Q~,
                                                 \eqno(16)
$$
describes a reaction front which moves with a constant speed
$$
S_\xi =  2^{-l}L \sqrt{\xi_{\rm split}}~,                  \eqno(17)
$$
and has a thickness 
$$
\delta_\xi\simeq 2^{-l}L/\sqrt{\xi_{\rm split}}~.          \eqno(18)
$$
As the front moves, it leaves behind cells marked with $\xi=1$.

The values of $\xi$ from (10) are used as the initial condition for
(14). The equation (14) then describes a reaction front 
which starts at places where $\xi>\xi_{\rm split}$, and propagates outwards 
normal to itself with the speed $S_\xi$. By integrating  (14), 
this front is advanced approximately $2-3$ computational cells.
According to the Huygens principle, the front curvature tends to decrease 
with time,  so that boundaries of regions marked for refinement
become smoother. Diffusion prevails over reaction in 
isolated areas with
 $\xi>\xi_{\rm split}$ if these areas have a size less than 
 $\simeq \delta_\xi$. These small regions do not trigger a reaction front,
and disappear under the refinement threshold (``extinguish'').
This reduces the numerical noise in selecting cells for refinement 
and unrefinement. To further reduce mesh trashing,
split cells are not joined if this produces an isolated leaf.

\bigskip\centerline{\bf 7. Numerical examples}

\bigskip
Many numerical tests 
including
advection, shock propagation and interaction in one, two and three 
dimensions were used to validate the algorithms.
Here we show some test examples
which illustrate adaptive mesh refinement aspects of
the algorithm's performance.

\bigskip\centerline{\it 7.1 Isolated shock wave passing through a
                         coarse-to-fine interface}

\bigskip
Figure 4 shows a computation of 
an isolated $M=10$ strong shock in an ideal gas with 
$\gamma=1.4$. In all three panels, the shock is moving from
left to right. The upper panel shows the shock profile at one time
during the shock propagation through a uniform mesh. The shock profile is
two cells wide, and oscillations free.
When the shock  passes from a fine to a coarse mesh (middle panel),
 the amplitude of disturbances generated 
is  less than 1\%. When passing from a coarse to a fine mesh (low panel),
 two types of disturbances,
acoustic and entropy waves, are generated at the $\simeq 5$\% amplitude.

A generation of spurious disturbances by a shock passing from a coarse to a
fine mesh is a known problem [2,4,16].
 An illustrative example for a $M=10$ shock in an ideal gas
with $\gamma=1.4$ is presented by Berger and Colella [2].
Berger and Colella argue that the disturbances are caused by 
the $O(1)$ errors always present in the numerical fluxes in the vicinity of 
a strong shock. The errors 
do not cancel each other if mesh spacing is changing. 
Formally, Berger and Colella's integration was second-order accurate at 
interfaces. In
 our case, the formal order of integration  at interfaces is
 first order. However, our example shows smaller disturbances, 
which could be attributed to the smaller refinement ratio (2:1 instead of 4:1).
The main conclusion from this test is that care must be taken not to 
let a strong shock enter
a more refined mesh. Strong shocks should be either refined 
all of the time, or not refined at all.

\bigskip\centerline{\it 7.2 Density discontinuity passing through a 
                       coarse-to-fine interface}

\bigskip
Figure  5 shows
the advection of a slab of high density, cold gas surrounded by a hot gas of
lower density. Initial conditions for this
problem are $P=0.01$ and $U=2$ everywhere, $\rho_{\rm cold} = 3$,
$\rho_{\rm hot}=1$. Initially, the slab is 20 cells wide.
The equation of state is 
that of an ideal gas with $\gamma = 1.4$ constant. 
Integration is done with $cfl = 0.7$.
The only result of passing a slab through coarse-to-fine interfaces
is changes in a number of cells representing contact discontinuities.
The pressure and velocity field (not shown) both remain constant
to machine accuracy. The  test shows that 
passing a discontinuity through coarse-to-fine interfaces does not generate
any disturbances.
However, if a 
discontinuity exits a refined area, the resolution is lost.
If it enters a refined area, its  width does not decrease,
and no gain in resolution is obtained. 

\bigskip\centerline{\it 7.3  Planar strong point explosion}

\bigskip
Figures 6 and 7 shows a one-dimensional (planar) strong point
 explosion in an ideal gas with  $\gamma=1.4$.
Initial conditions are $\rho_0 = 1$, $P_0 = 10^{-3}$ 
everywhere. The explosion energy $E = 2.5\times 10^9$ is put in a single cell
by increasing the pressure in that cell to
$P = P_0 + (\gamma -1) E/\Delta$, where $\Delta$ is the cell size.
Figure 6 shows the explosion computed on a uniform mesh with 
$\Delta={1\over 64}$.
The exact solution to the problem is shown for comparison [29].
The solutions agree  near the center. 
The difficulty is to reproduce the solution near the shock.

Figure 7  shows the same explosion computed with eight levels of 
refinement,  $l = 5\div 12$. Two refinement indicators were used,
one for shocks (11) and another one for the pressure
gradient (13). The refinement criteria were
$\xi_{\rm split} = 0.5$ and $\xi_{\rm join}=0.01$.
 In this case, the flow behind the shock has been resolved. 
The number of cells used in the simulation was $\simeq 250$, and only a few 
($< 20$) located at the highest level $l=12$. 
A uniform $4096$ mesh would be required to get the same resolution. 
Small disturbances at fine--coarse interfaces are
seen at the pressure and velocity plots. Since the gradients in the
solution are large everywhere, these disturbances may be caused
by the same inexact cancelling of numerical errors in fluxes 
at the fine--coarse interfaces  discussed in [2] and in Section 7.1 above,
 or they may be the result of decreasing
the order of the accuracy of integration at interfaces, or both.
Which effect is important, and whether it is worthwhile to
maintain  formal second-order accuracy at interfaces, requires further study.

\bigskip\centerline{\it 7.4  Cylindrical strong point explosion in a box}

\bigskip
Figures 8 and 9 show 
a cylindrical strong point explosion in a square box, and 
illustrate how the adaptive mesh algorithm treats a 
very complicated, rapidly
evolving flow with multiple shocks, contact discontinuities, and vortices.
Initial conditions in the box are $\rho_0 = P_0 = 1$.
The equation of state is that of an ideal gas with $\gamma = 1.4$.
The explosion energy $E=10^5$ is put in a single cell of size $1\over 1024$
located at $x=0.35$, $y=0.2$. 
The computation was done with six levels of refinement, $l=5\div 10$,
using shock and pressure gradient refinement criteria with
$\xi_{\rm split} = 0.5$ and $\xi_{\rm join}=0.05$. The Courant number, $cfl$,
 is $0.7$. The equivalent uniform resolution would require a $1024\times 1024$
grid or $2^{20} = 1,048,576$ cells.

A cylindrical blast wave propagates from the
 explosion point, and first hits the lower, then the left, right and upper 
walls of the box. Reflections of the primary shock begin as regular 
reflections, but later transform into Mach reflections. 
The sound speed is very high in the material behind the primary shock.
As a result, reflected shocks quickly engulf the inside of the hot bubble 
created by the initial blast, catch up
with the primary shock, and interact with walls and with each other. 
Figures 8a (step 200) and 8b (step 300) show 
the interaction of the primary shock
 with the lower wall.
In Figure 8c (step 400)
 this interaction continues, but the shock is also reflecting
off the left wall. Reflections of the primary shock off the walls are irregular.
In the next Figure 8d (step 500), a part of the primary shock
 is still interacting with the 
upper and right walls. In Figure 8e (step 600), the primary shock disappeared,
 and the collision of
the two reflected shocks just occurred near the upper right corner of the
box. Figure 8f shows the pressure contours for step 600.
 The mesh for the two time steps, 400 and 600, is shown
in Figure 9. Figures 8 and 9 show a complicated pattern of secondary 
shocks resolved by the  mesh refinement  algorithm.

\bigskip\centerline{\sl 7.5 Shock -- bubble interaction in three dimensions}

\bigskip
The interaction of shock waves with fluid inhomogeneities
is an important mechanism of vorticity generation [30,31].
The interaction of a planar shock with 
a low-density spherical bubble of gas is one of the four
 representative cases  studied theoretically in [31]. 
It is  assumed in [31] that the 
gas in the bubble is in pressure equilibrium with the ambient 
gas, but has higher temperature. This is different from the
conditions of the experiments [30] where the density contrast
was created by using gases with different molecular 
weight. Formulation of the problem in [31] is highly relevant to
the shock--generated vorticity in flames.

The computational setting for the shock -- bubble problem is shown in
 Figure 10. A spherical bubble of hot, low density gas is located
inside a rectangular, solid wall tube of length 1 and cross-section 
${1\over 2}\times {1\over 2}$. 
The computational domain of size
$L_x \times L_y\times L_z = 1 \times {1\over 4} \times {1\over 4}$
represents a quarter of the tube cross-section assuming symmetry 
with respect to the middle planes $y={1\over 4}$ and $z={1\over 4}$.
The radius of the  bubble is $R_b = {1\over 8}$. The bubble is centered
at a distance $x_b = {1\over 4}$ from the tube entrance.
The bubble and the background gas are in the
pressure equilibrium at a common pressure $P_0 = 1$. The background
density is $\rho_0 = 1$ and the bubble density is $\rho_b=0.166$.
 The equation of state is that of an ideal gas with
constant $\gamma = 1.4$. 
A shock wave with the Mach number $M=1.25$ hits the bubble from
the left. Boundary conditions are inflow on the tube entrance, and outflow on 
the tube exit. Simulations are performed using five levels of refinement 
$l=5\div 9$. The maximum equivalent uniform resolution would correspond to a
$512\times 128\times 128$ grid, and would require $2^{23} = 8,388,608$ cells
of size  $1\over 512$.
The three-dimensional resolution in this test case is 20\% higher than that
 used in two-dimensional simulations [31]. 
Four refinement indicators were used:
for shocks (15), for contact discontinuities (16) and for density and pressure
gradients (17). The refinement criteria are
$\xi_{\rm split} = 0.5$ and $\xi_{\rm join}=0.05$. The Courant number
$cfl=0.7$ was used.

Figure 11 shows the density contours and the mesh in the XY plane 
$z={1\over 4}$ 
after 20 and 70 global time steps of integration. Figure 12 shows
the density contour and the mesh for the same time steps in the 
perpendicular YZ-planes $x=0.35$ and $x=0.6$, respectively.
 The shock compresses the bubble,
 and vorticity is generated due to
misalignment of the pressure and density gradients during the 
interaction. Vorticity is generated mainly at the bubble surface
where transmitted shocks are refracted.
Rotational motions generated as a result of shock refraction
 further deform the bubble, and 
eventually transform it into the rotating ring. The results of the
simulation presented here are in a good general 
agreement with the results of two-dimensional axially-symmetric simulations
[31]. 
Three-dimensional simulations, however,
allow assumption of axial symmetry to be relaxed. Figure 12 shows
the development of azimuthal perturbations on the surface of the
bubble as time progresses. These perturbations are initially caused by the 
the Rayleigh-Taylor instability
driven by the deceleration of material between the
bubble and the spherical transmitted shock. These  perturbations
are further  subjected to the Kelvin-Helmholtz instability which develops
due to  the presence of the velocity component tangent to the bubble surface.
The tangential component has been
 generated during the shock refraction at the bubble surface.
A discussion of instabilities developing during the shock-bubble
interaction is out of scope of this paper, and will be presented elsewhere.

\bigskip\centerline{\bf 8. Algorithm performance} 

\bigskip
The algorithm consists of four parts: 
(1) FTT subroutines for tree modifications,
dumping, performing global parallel operations, finding
neighbors, parents and children, etc.; (2) Fluid dynamics subroutines;
 (3) Refinement subroutines; (4) Service subroutines: driver, print, plot,
etc. The code is written entirely in Fortran 77. The line count for
 each part is given in Table 1 (comments included). 
The critical FTT part on which the rest of the algorithm 
is built contains less than 800 lines. 

Fourth column of Table 1 shows the breakdown of computer time for
different parts of the algorithm compiled using a Sun f77 compiler with -O5 
optimization, and run on a  Sun workstation  under Solaris operating system.
The figures show that the FTT part of the code consumes a relatively small
fraction of time.
The fifth column gives the time breakdown for the same code compiled
 using Cray cf77 compiler with -Zu parallel option
enabled, and run on a single Cray J90 processor. On a vector machine,
 the relative overhead for the FTT is noticeably larger, 
and comprises about 1/3 of the
total time. This is because FTT uses indirect addressing
which slows vector operations. 	In the 
fluid dynamics part of the code, work is done by looping over contigous
blocks of memory, and all loops are effectively vectorized. The J90 computers
lack hardware gather/scatter capabilities, have only two
paths between the vector processor unit and the memory, and have a rather long
memory latency. The FTT part will run more effectively on Cray C90 or
 T90 computers. On a single processor of Cray J90, the computer time to 
advance one cell one  sub-step is 30 $\mu$sec per dimension.	
In a multi-processor mode, the code used up to  10 processors
of a 16 processor J90 in a batch environment.

\bigskip
\centerline{Table 1}
\medskip
\settabs\+\noindent
  &***&***********************&*****************&********************&\cr
\hrule\smallskip
\+&      & algorithm's part      & lines     & time, \% (SUN) & time, \% (J90)  \cr
\smallskip\hrule\smallskip
\+&   & FTT                   & 800       &  11       &  35         \cr
\+&     & Hydro                 & 1485      &  83       &  61         \cr
\+&     & Refinement            & 601       &  3        &  2          \cr
\+&     & Services              & 626       &  3        &  2          \cr
\+&     & Total                 & 3512      &  100      &  100        \cr
\smallskip\hrule

\bigskip\noindent
For the cylindrical strong point explosion,
 and for the shock--bubble interaction tests,
Figures 13 and 14 show the ratio of the number of computational cells 
actually used in the simulation to the  number of cells required to achieve an
equivalent uniform resolution. The ratio varies with time. For the cylindrical
shock test, for example, it is almost zero in the beginning, when the
shocked material occupies only a small fraction of the computational
domain. Then the ratio grows with time almost linearly because fine cells
tend to concentrate around the surface of the shock only.
At the end of the simulation, the ratio reaches its highest value
$\simeq 0.1$. At this time the entire computational domain is filled with
shocks, and the regions of constant flow have disappeared.
On average, the ratios for both simulations are $\simeq 0.05-0.07$
which represents a factor of $\simeq 15$ savings in memory and computer time
compared to a brute force uniform grid computations. Here we have taken 
into account that the FTT overhead would be absent in grid-based computations.

\bigskip\centerline{\bf 9. Conclusions}

\bigskip
A fully threaded tree structure (FTT) for adaptive
refinement of regular meshes has been described. 
The FTT allows all operations and modifications of the tree to be
performed in parallel, which makes this structure well suited for 
the use on vector and parallel computers. 
A filtering algorithm  has been described for removing high-frequency noise in
mesh refinement.

The FTT has been applied to the integration of 
the Euler equations of fluid dynamics.
Time stepping, and mesh refinement algorithms specific to
the integration of the Euler equations were described.
The integration in time is done using different time steps at
different tree levels. The integration and mesh refinement are interleaved
to avoid creation of buffer layers of fine mesh ahead of
moving shocks and other discontinuities. 
The time stepping algorithm described can utilize any method of 
evaluating numerical fluxes which is at least second
order accurate in space and time on a uniform mesh.

The FTT performance on a Sun workstation, and on
a shared memory Cray J90 computer is reported.
The FTT has  low memory and computer time overheads, 2 words
and $\le 10-40$\% of computer time per computational cell, respectively.
A factor $\simeq 15$ savings  in memory and computer time compared to 
equivalent uniform-grid computations have been achieved 
in two- and three-dimensional test simulations presented in the paper. 
Three-dimensional simulations of a shock wave interacting with a spherical
bubble of light gas have been performed that 
 shown the development of azimuthal instabilities
at the bubble surface.

There are many uses of FTT other than the integration of the Euler equations,
such as the solution of parabolic and
elliptical partial differential equations, pattern recognition,
computer graphics, particle dynamics, etc.
An application of the adaptive refinement tree to
 dark matter  cosmological simulations has been recently described in [25].
An application of FTT to 
reaction--diffusion equations will be described elsewhere. 

\bigskip
Parts of the work were supported by the NSF grant AST94-17083, 
NASA grant NAG52888, by 
the DARPA Applied and Computational Mathematics Program,
and by the 
Office of Naval Research.

The author is grateful to Almadena Chtchelkanova 
(Berkeley Research Associates,  Inc.)  for her advices on various 
data structures and algorithms, to Sandy Landsberg for references, to
Elaine Oran and Jay Boris for useful discussions, and to Elaine Oran
 for reading the manuscript.
Discussions with Andrey Kravtsov 
and Anatoly Klypin (NMSU)  during the final stage of code development 
resulted in  several improvements of the FTT structure.
\vfill\eject

\centerline{\bf Literature}

\item{1.} 
M. J. Berger and J. Oliger, {\sl J. Comput. Phys.} {\bf 53}, 484 (1984)

\item{2.}
M. J. Berger and P.  Colella, {\sl J. Comput. Phys.} {\bf 82}, 64 (1989)

\item{3.}
M. J. Berger and R. J. Leveque, {\sl AIAA paper 89-1930-CP} (1989)

\item{4.}
J.J. Quirk, {\sl PhD Thesis, Cranfield Institute of Technology, U.K} (1991)

\item{5.}
J. J. Quirk, {\sl Computers Fluids} {\bf 23}, 125 (1994)

\item{6.}
J. J. Quirk and S. Karni, {\sl J. Fluid Mech.} {\bf 318}, 129 (1996)

\item{7.}
R. B. Pembert, J. B.  Bell, P. Colella, W. Y. Crutchfield, and
M. L. Welcome, {\sl AIAA paper 93-3385-CP} (1993)

\item{8.}
J. {Bell, M. J. Berger, J.  Saltzman, and M. Welcome,
        {\sl SIAM J. Sci. Comput.} {\bf 15}, 127 (1994)

\item{9.}
M. Ruffert, {\sl Astronomy \& Astrophys.} {\bf 265}, 82 (1992)

\item{10.}
M. Ruffert and W. D. Arnett, {\sl Astrophys. J.}{\bf 427}, 351 (1994)

\item{11.}
R. I. Klein, C. F.  McKee, and P. Colella, {\sl Astrophys. J.} {\bf 420},
                                                               213 (1994)

\item{12.}
G. G. Duncan and P. A. Hughes, {\sl Astrophys. J. Lett.} {\bf 436}, L119 (1994)

\item{13.}
D. P. Young, R. G. Melvin, M. B. Bieterman, F. T. Johnson, S. S. Samant, and
J. E. Bussoletti, {\sl J. Comput. Phys.} {\bf 92}, 1 (1991)

\item{14.}
D. D. Zeeuw  and K. G. Powell, {\sl J. Comput. Phys.} {\bf 104}, 56 (1993)

\item{15.}
M. G. Edwards, J. T. Oden, and L. Demkowicz,
        {\sl SIAM J. Sci. Comput.} {\bf 14}, 185 (1993)

\item{16.}
W. J. Coirier, {\sl PhD Thesis, Unuiversity of Michigan} (1994)

\item{17.}
W. J. Coirier and  K. G.  Powell, {\sl J. Comput. Phys.} {\bf 117}, 121 (1995)

\item{18.}
M. J. Berger and  J. E. Melton, {\sl preprint, to appear in 
                                             Proc. 5th Intl. Conf. Hyp. Prob.,
                                             Stonybrook, NY} (1994)
\item{19.}
M. J. Aftosmis, J. E.  Melton and M. J.  Berger, {\sl AIAA paper 95-1725-CP} 
(1995)

\item{20.}
J. E. Melton, M. J.  Berger, M. J. Aftosmis, and M. D.  Wong,
        {\sl AIAA paper 95-0853} (1995)

\item{21.}
Yu-Liang Chiang, B. van Leer, and K. G. Powell, {\sl AIAA paper 92-0433} (1992)

\item{22.}
S. A. Bayyuk, K. G. Powell, and B. van Leer, {\sl AIAA paper 93-3391-CP} (1993)

\item{23.}
S. A. Bayyuk, K. G. Powell, and B. van Leer, {\sl Proceedings, The First
AFOSR Conference on Dynamic Motion CFD, New Brunswick, New Jersey, 
3-5 June 1996}

\item{24.}
N. Dale and H. M. Walker, {\sl Abstract Data Types: Specifications, 
      Implementations, and Applications} (Jones and Bartlett, Boston, 1996)

\item{25.}
A. V. Kravtsov, A. A. Klypin, and A. M. Khokhlov, 
{\sl Ap.J. Suppl}, 1996, submitted, preprint {\tt astro-ph/9701195}.

\item{26.}
P. Colella and H. M. Glaz, {\sl J. Comput. Phys.} {\bf 59}, 264 (1985)

\item{27.} B. van Leer, {\sl J. Comput. Phys.} {\bf 32}, 101 (1979) 

\item{28.}
P. Colella and P. R.  Woodward, {\sl J. Comput. Phys.} {\bf 54}, 174 (1984)

\item{29.}
V. P. Korobeinikov, {\sl Zadachi Teorii Tochechnogo Vzryiva} (Moskva, Nauka,
1985) p. 78

\item{30.}
J.-F. Haas and B. Sturtevant, {\sl J. Fluid Mech.} {\bf 181}, 41 (1987)

\item{31.}
J. M. Picone and J. P. Boris, {\sl J. Fluid Mech.} {\bf 188}, 23 (1988)

\item{32.}
P. R. Woodward and P. Colella, {\sl J. Comput. Phys.} {\bf 54}, 115 (1984)

\item{33.}
D. H. Porter, P. R. Woodward, W. Yang, and Qi Mei,  in
{\sl Nonlinear Astrophysical Fluid Dynamics} 
eds.\ {Buchler, R.J., and Gottesman, S.T.}, New
York Academy of Sciences,  New York, 1990, pp.\ 234-258

\item{34.}
E. S. Oran and J. P. Boris {\sl Computers in Physics} {\bf 7}, 523 (1993)

\item{35.}
E. S. Oran and J. P. Boris {\sl Numerical Simulations of Reactive Flows},
  Elsevier Science Publishing Co., Inc., New York (1987).
\vfill\eject

\centerline{\bf Figure captions}

\bigskip\noindent
Fig. 1 -- Relationship between cells for a one-dimensional,
binary, fully threaded tree. 
Each cell is represented by a horizontal barred line.
The length of a line corresponds to the geometrical size of a cell. 
Cell 1 is the root of the tree, representing the
entire computational domain. Cells 1 and 2 are split cells.
Cells 3, 4 and 5 are leaves. Pointers to children and parents
 are indicated by  straight lines without arrows. Pointers to
neighbors are indicated by arrows. 

\bigskip\noindent
Fig. 2 -- Relationship between the cells and octs in a fully threaded
tree. Pointers from octs to cells and from cells to octs
are indicated by arrows. 

\bigskip\noindent
Fig. 3 -- An illustration of flux evaluation at two different levels of the 
         tree. Fluxes at interfaces between fine cells (1--3 and 2--4),
          and fluxes between
         fine and coarse cells (3--5 and 4--5)
          are evaluated twice as often as
         fluxes between coarse cells (5--6).
         Four fluxes from the fine side
         (3--5 and 4--5 computed at two different times), and
         one flux (5--6) from the coarse side contribute to changes of the
         state vector of cell 5 during a coarse time step.

\bigskip\noindent
Fig. 4 -- An isolated shock wave with a Mach number $M=10$ in an ideal gas with
$\gamma = 1.4$; computed with $cfl = 0.7$. 
Density is shown by dots whose
position corresponds to centers of computation cells.
In all panels, the shock is moving from left to right.
 Top panel -- shock on a uniform mesh.
Middle panel -- shock after passing from fine to coarse mesh.
Bottom panel -- shock  after passing from coarse to fine mesh. 
Coarse-to-fine interface is located near $x\simeq 0.11$.
Two disturbances generated during
the passage of the coarse-to-fine interface are seen on the bottom panel.

\bigskip\noindent
Fig. 5 -- Advection of a high density, cold slab of gas with $\rho_{\rm high}=3$ surrounded by a low  density hot gas with $\rho_{\rm low}=1$.
Pressure is $P=0.01$, and velocity is $U=2$ everywhere.
An ideal gas EOS with $\gamma = 1.4$; computed with $cfl = 0.7$. 
Advection is taking place from left to right.
A coarse-to-fine interface is located at $x\simeq 0.38$ (dashed line).
Density is shown by dots whose
position corresponds to centers of computation cells. Time-step number
is indicated at the top of each density profile.

\bigskip\noindent
Fig. 6 -- Strong point explosion in an ideal gas $\gamma=1.4$ in planar 
geometry. The initial pressure is  $P_0 = 10^{-3}$, initial density is
$\rho_0=1$, and the explosion energy is $E=2.5\times 10^8$. Computed on
a uniform mesh using $cfl=0.7$. Plots of density, pressure, velocity, and
cell level in the tree are shown for three different moments of time
(1) -- $1.57\times 10^{-6}$, (2) -- $3.49\times 10^{-6}$, 
and (3) -- $5.82\times 10^{-6}$. 
The variables are shown by dashed lines
and by dots whose
position corresponds to centers of computation cells.
Solid lines -- the exact solution.

\bigskip\noindent
Fig. 7 -- Strong point explosion in an ideal gas $\gamma=1.4$ in planar 
geometry. The initial pressure is  $P_0 = 10^{-3}$, initial density is
$\rho_0=1$, and the explosion energy is $E=2.5\times 10^8$. Computed using
$cfl=0.7$ and eight levels of mesh refinement.
Plots of density, pressure, velocity, and
cell level in the tree are shown for three different moments of time
(1) -- $1.15\times 10^{-6}$, (2) -- $3.02\times 10^{-6}$, 
and (3) -- $6.07\times 10^{-6}$. 
The variables are shown by dots whose
position corresponds to centers of computation cells.
Solid lines -- the exact solution.

\bigskip\noindent
Fig. 8 -- Cylindrical strong point explosion in a square box with solid walls.
Computed with levels of refinement $l=5\div 10$.
Density is shown for time steps: (a) -- 200, (b) - 300, (c) -- 400,
(d) -- 500, and (e) -- 600; (f) -- pressure for time step 600.
Density contours are $\Delta \rho = 0.1$ beginning with $\rho_{\rm min}=0.05$.
Pressure contours are $\Delta P = 10^3$ beginning with $P_0 = 0$.
Coordinates are in units of the finest cell size $\Delta(10) = {1\over 1024}$.

\bigskip\noindent
Fig. 9 -- Same as Fig. 8 but shows the computational mesh: (a) -- step 400,
and (b) -- step 600.

\bigskip\noindent
Fig. 10 -- Computational setting of the shock wave -- spherical bubble 
interaction problem. Depicts a square tube of size
 $1\times {1\over 2}\times {1\over 2}$. The computational domain is a
quarter of the tube cross-section shown by thick lines and marked by the
numbers 1 to 8, assuming symmetry with respect to the XZ plane $y={1\over 4}$
and XY plane $z={1\over 4}$.

\bigskip\noindent
Fig. 11 -- Shock wave -- spherical bubble interaction.
Density contours $\Delta \rho = 0.05$ and the mesh in the
XY plane 1-2-5-6 (see Fig. 10) for time steps 20 and 70.

\bigskip\noindent
Fig. 12 -- Shock wave -- spherical bubble interaction.
Density contours $\Delta \rho = 0.05$ and the mesh in the
YZ plane for time steps 20 and 70. The plane coordinates are 
$x=0.35$ (step 20) and $x= 0.6$ (step 70).

\bigskip\noindent
Fig. 13 -- The ratio of actually used computational cells to the number of 
uniform grid cells for various time time steps of the cylindrical strong
point explosion problem (Section 7.4).

\bigskip\noindent
Fig. 14 -- The ratio of actually used computational cells to the number of 
uniform grid cells for various time time steps of the shock -- spherical bubble
interaction problem (Section 7.5).

\vfill\eject
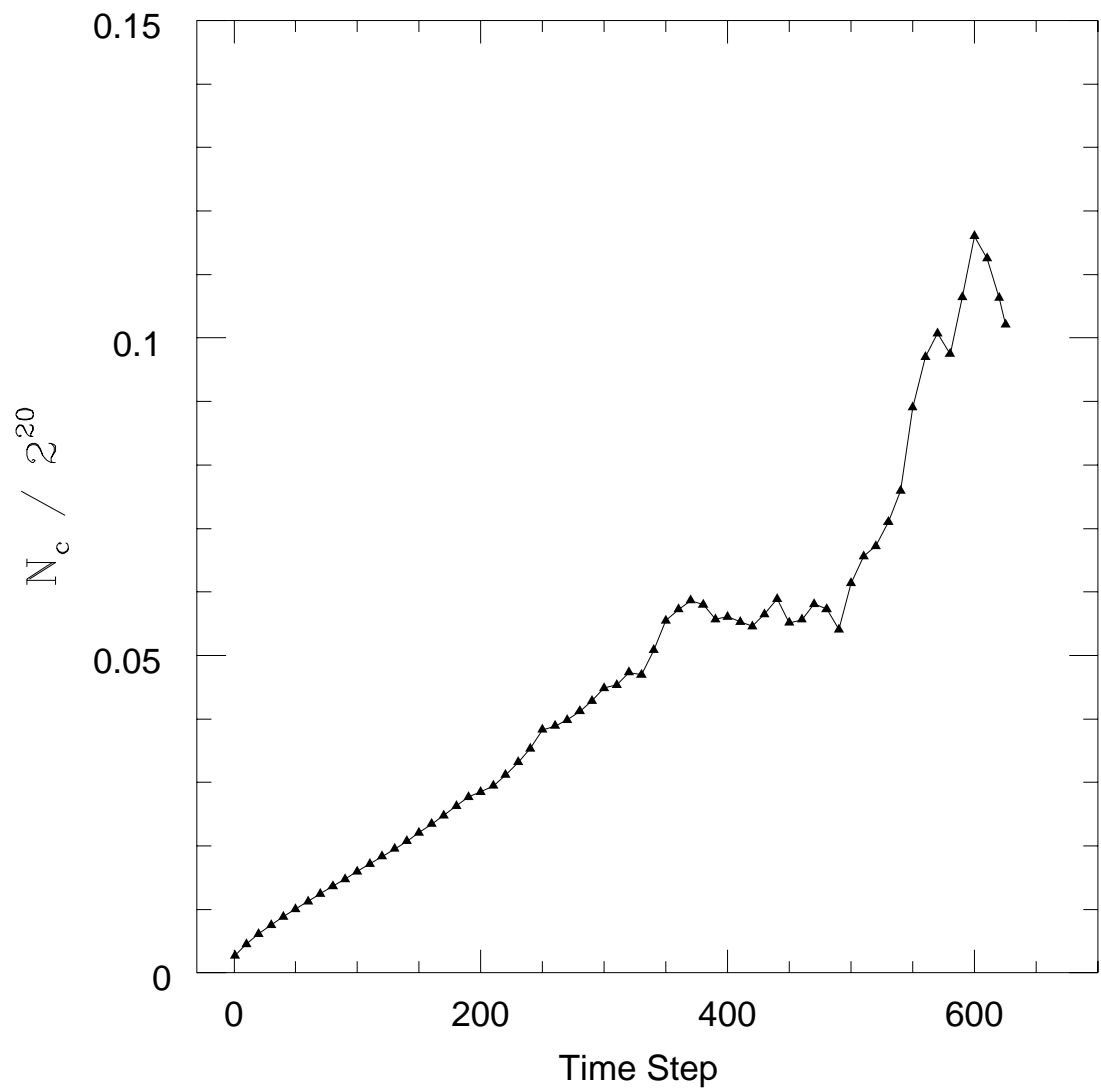
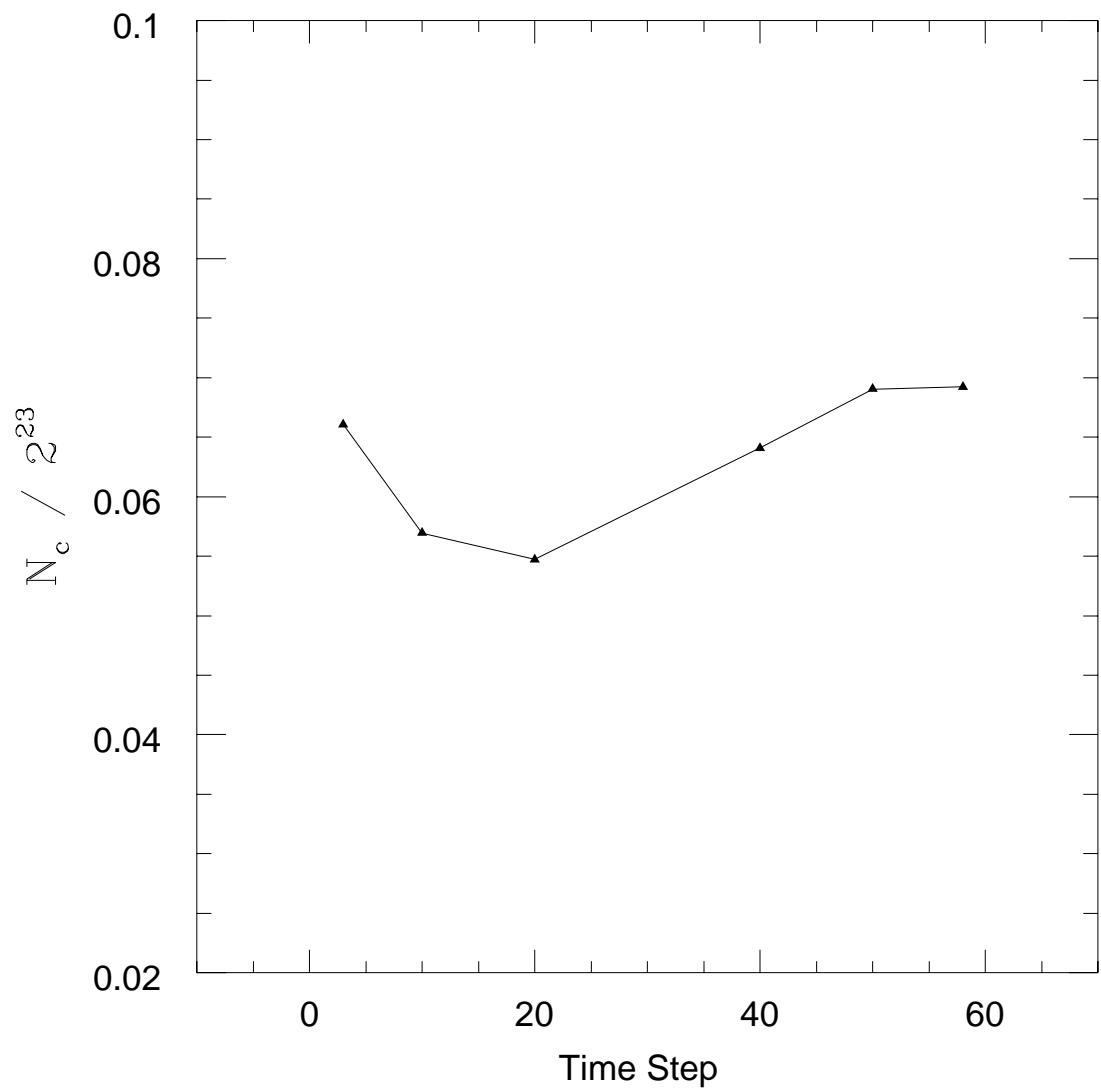
\end